\documentclass[]{spie}  

\usepackage[]{graphicx}
\usepackage{amsmath}

\newcommand{\reffig}[1]{Figure \ref{#1}}
\newcommand{\refnum}[1]{Ref.~\citenum{#1}}

\title{Automated Alignment and On-Sky Performance of the Gemini Planet Imager Coronagraph}

\author{Dmitry Savransky\supit{a}, Sandrine J. Thomas\supit{b}, Lisa A. Poyneer\supit{c},   Jennifer Dunn\supit{d}, Bruce A. Macintosh\supit{e}, Naru Sadakuni\supit{f}, Daren Dillon\supit{g}, Stephen J. Goodsell\supit{f}, Markus Hartung\supit{f}, Pascale Hibon\supit{f}, Fredrik Rantakyr{\"o}\supit{f}, Andrew Cardwell\supit{f}, Andrew Serio\supit{f} with the GPI team
\skiplinehalf
\supit{a}Sibley School of Mechanical and Aerospace Engineering, Cornell University, Ithaca, NY USA\\
\supit{b}NASA Ames Research Center,  Mountain View, CA USA\\
\supit{c}Lawrence Livermore National Laboratory, Livermore, CA USA\\
\supit{d}National Research Council of Canada Herzberg, Victoria, B.C. Canada\\
\supit{e}Kavli Institute for Particle Astrophysics and Cosmology, Stanford University, Stanford, CA USA\\
\supit{f}Gemini Observatory, La Serena, Chile\\
\supit{g} University of California Observatories/Lick Observatory, University of California, Santa Cruz,  Santa Cruz, CA USA
}

\authorinfo{Send correspondence to Dmitry Savransky 
\texttt{ds264@cornell.edu}}

\begin{document} 
\maketitle 

\begin{abstract}
The Gemini Planet Imager (GPI) is a next-generation, facility instrument currently being commissioned at the Gemini South observatory.  GPI combines an extreme adaptive optics system and integral field spectrograph (IFS) with an apodized-pupil Lyot coronagraph (APLC) producing an unprecedented capability for directly imaging and spectroscopically characterizing extrasolar planets.  GPI's operating goal of $10^{-7}$ contrast requires very precise alignments between the various elements of the coronagraph (two pupil masks and one focal plane mask) and active control of the beam path throughout the instrument.  Here, we describe the techniques used to automatically align GPI and maintain the alignment throughout the course of science observations.  We discuss the particular challenges of maintaining precision alignments on a Cassegrain mounted instrument and strategies that we have developed that allow GPI to achieve high contrast even in poor seeing conditions.
\end{abstract}


\section{INTRODUCTION} \label{sec:intro} 

The development and refinement of direct imaging is extremely important to advancing our understanding of the formation and evolution of exoplanetary systems. While indirect detection methods have proven very successful in discovering exoplanets, they are dependent on collecting multiple orbits of data and are thus biased towards short-period planets.  Direct imaging, on the other hand, is biased towards planets on larger orbits, making it highly complementary to the indirect methods.  Together, these techniques can significantly advance our understanding of planet formation and evolution by providing a sample of planets at all orbital scales.  Additionally, direct imaging provides the most straightforward way of getting planet spectra, which are invaluable to the study of planetary and atmospheric compositions and can serve as proxies for planet mass measurements \cite{barman2011clouds}.

The Gemini Planet Imager (GPI) is a facility instrument for the Gemini South observatory, designed specifically to directly image young, self-luminous, giant extrasolar planets.  Delivered to the observatory in late 2013  after extensive integration and testing at the University of California, Santa Cruz \cite{hartung2013final}, GPI saw its first light in November of that year, and has been undergoing commissioning and verification, and performing some early science observations since then\cite{macintosh2014first}.   GPI is scheduled to begin an 890 hour exoplanet survey of 600 nearby, young stars in the second half of 2014  with an expected yield of 25 to 50 new, directly imaged exoplanets \cite{mcbride2011experimental,savransky2013campaign}. 

GPI is significantly more complex than previous exoplanet imaging instruments, with much tighter tolerances on alignments and performance which lead to a particular set of challenges.  Furthermore, as GPI is Cassegrain-mounted and moves with the telescope, it experiences an ever changing gravity vector and must also be completely self-contained, with no manual intervention in its internal alignments.  GPI also has multiple possible internal configurations, all of which must be automatically accessible via remote control, with a high degree of repeatability. Finally, since GPI is a facility instrument and will be operated in queue-scheduled mode, it must be able to automatically carry out all of its tasks with good fault-tolerance and error handling.

The automation of GPI's operations is discussed in detail in \refnum{dunnthis}, while the integration of GPI with the Gemini South observatory is presented in \refnum{rantakyrothis}.  In this paper, we will focus on the particular challenges of automating the internal alignment of GPI's reconfigurable diffraction control system, and will present some on-sky results from GPI's recent first light runs. In \S\ref{sec:overview} we provide an overview of GPI's internal optics and the relevant portions of the instrument design, \S\ref{sec:coron} details the methods used for internal alignments, and \S\ref{sec:onsky} presents results from on-sky operations.

\section{GPI OVERVIEW}\label{sec:overview}
\begin{figure}[ht]
 \begin{center}
   \includegraphics[width=0.85\textwidth]{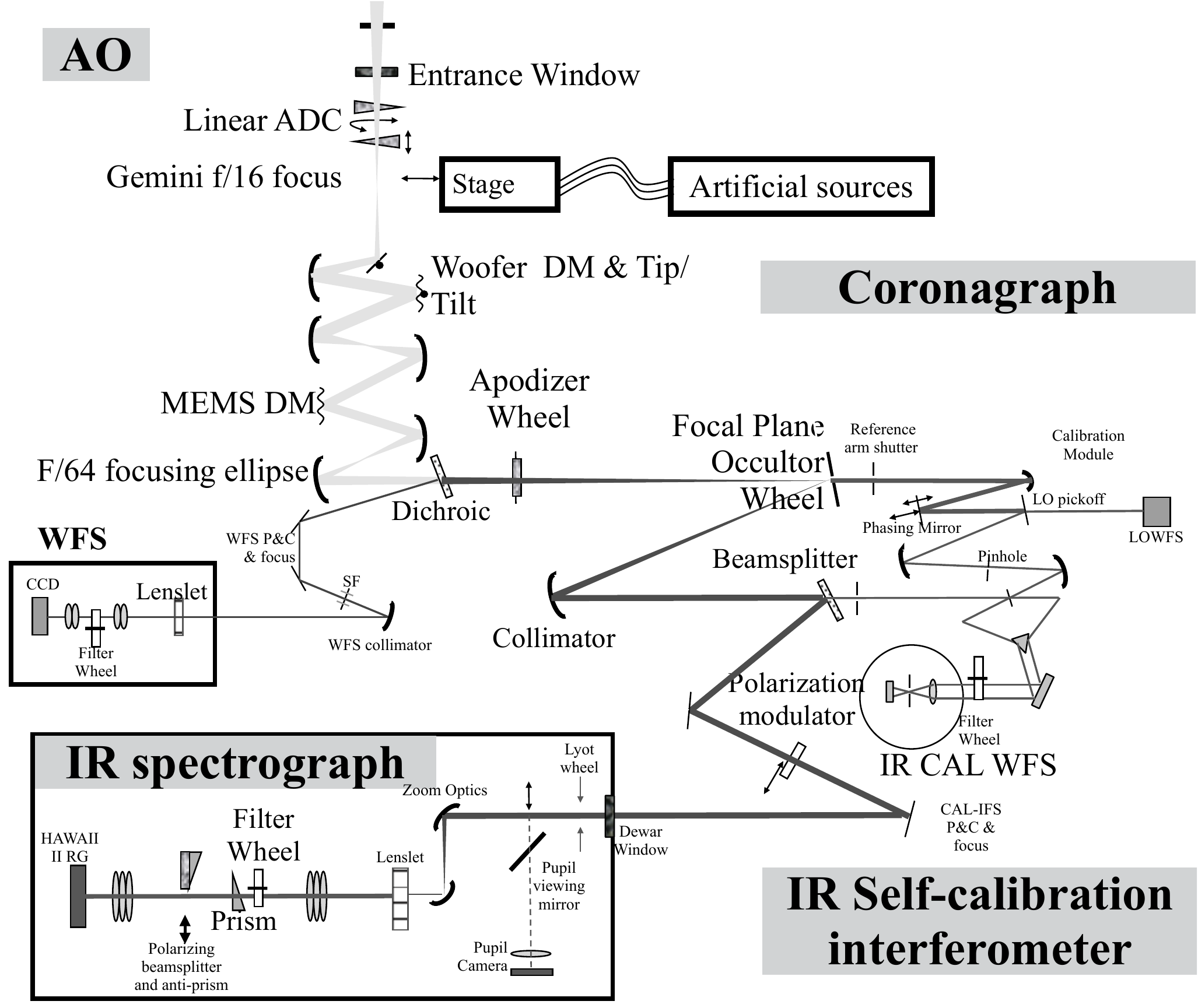} 
 \end{center}
 \caption[]{ \label{fig:gpi_lightpath} GPI light path. \cite{macintosh2008gemini} The instrument is composed of three subsystems - the Adaptive Optics system, the Calibration Interferometer, and the Integral Field Spectrograph science instrument, with components of the diffraction control system (an Apodized-Pupil Lyot Coronagraph) distributed throughout.  The coronagraph is composed of an apodizer in the final pupil plane of the AO system, a Focal Plane Mask in the Calibration system, and a Lyot stop in the entrance pupil of the science instrument.}
 \end{figure}
 
GPI's architecture and design have been extensively discussed (see, in particular, Refs.~\citenum{macintosh2008gemini,macintosh2012gemini,macintosh2014first}).  Here, we will present only a brief description of the instrument and discuss those components that are relevant to this work.  GPI consists of three main subsystems and a diffraction control system, whose components are spread throughout the instrument (illustrated in \reffig{fig:gpi_lightpath}).  Light from the telescope pupil, or an internal calibration source called the Artificial Source Unit (ASU), first passes through the Adaptive Optics (AO) system, which corrects phase errors introduced by the turbulent atmosphere and by internal surface errors.  The AO system uses a MEMS deformable mirror (DM), a piezo-electric DM and  a tip-tilt stage for control, operating in closed loop with a Shack-Hartmann wavefront sensor (WFS) consisting of a visible light  CCD behind a 43x43 lenslet array.  A variable-size spatial filter is included in this leg of the beam path to remove aliasing errors\cite{poyneer2004spatially} and an input fold mirror is used for initial steering of the beam, changing both the pointing and centering of the beam on the AO WFS.  The remaining IR light passes through a pupil plane apodizer,which is the first component of the diffraction control system---an apodized-pupil lyot coronagraph (APLC)\cite{soummer2004apodized}---and continues to the calibration interferometer (CAL).  The CAL system interferes on-axis light passing through the focal plane occulting mask (FPM; the second component of the APLC) with a portion of the reflected off-axis light in order to reconstruct the post-coronagraph wavefront and generate reference centroid offsets that are sent to the AO control loop\cite{wallace2008post}. The remaining IR light enters the science instrument, an integral field spectrograph (IFS)\cite{larkinthis}, where it is passed through a pupil plane Lyot stop (the final APLC component).  
 
  \begin{figure}[ht]
 \begin{center}
   \includegraphics[width=1\textwidth]{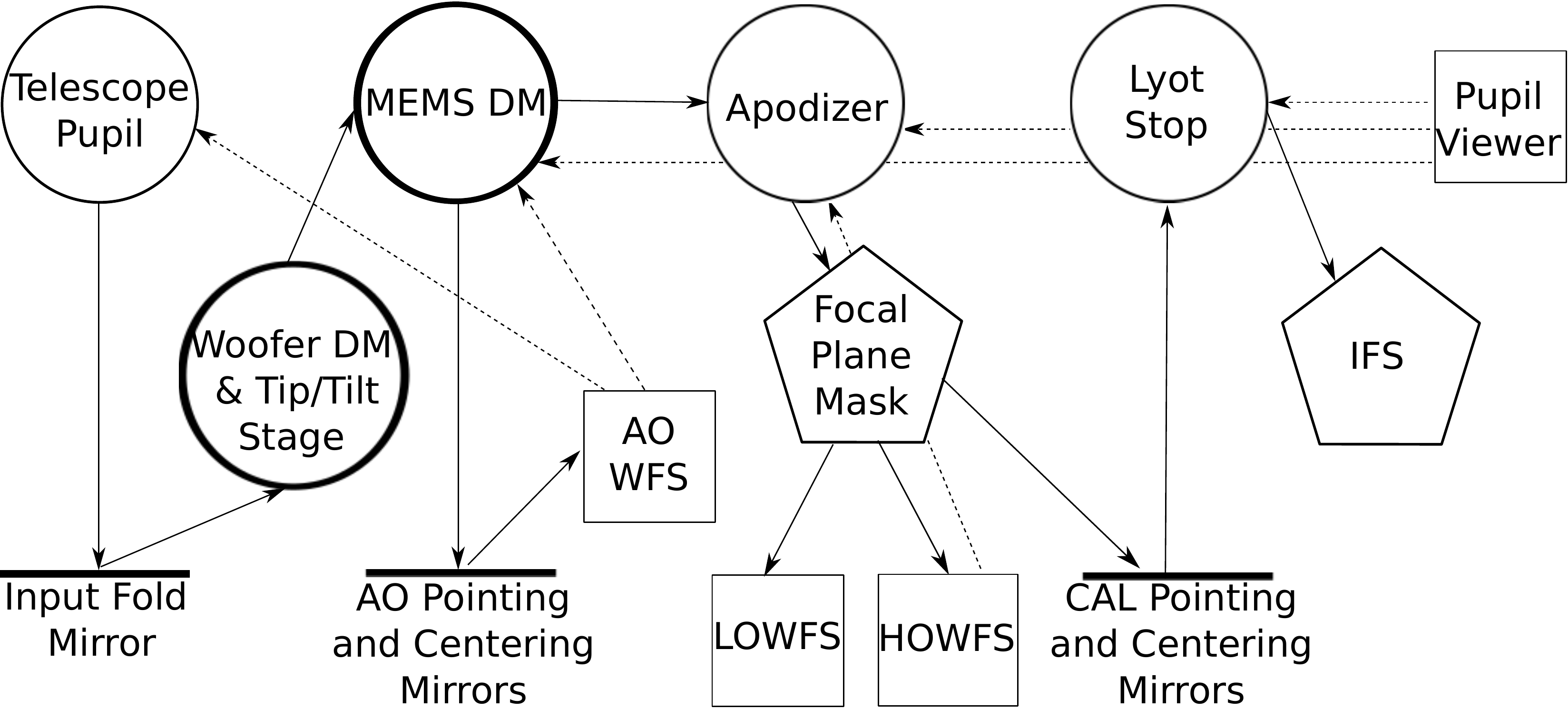}
 \end{center} 
 \caption[]{ \label{fig:gpi_schematic} GPI Schematic.  Circles represent pupil planes, pentagons are focal planes, squares are wavefront sensors and detectors.  Thick lines represent controllable surfaces such as moveable mirrors and stages, solid lines with arrows represent the light path, and dashed lines represent imaging of pupil planes by various detectors for alignment purposes.  Note that this is a schematic view only: multiple optics, including beam splitters and pupil re-imagers, are not shown here.  The four pupil planes at the top of the schematic must all be aligned to a common plane, with tolerances given in Table \ref{tbl:tolerances}.\cite{savransky2013computer}}
 \end{figure}

\begin{table}[ht]
\caption{\label{tbl:tolerances} GPI pupil alignment tolerances}
\begin{center}
\begin{tabular}{cp{1in}c c}
\hline
Alignment & Tolerance$^a$\ & Measurement & Achieved$^b$\\ 
\hline
Telescope Pupil to MEMS &  0.5\% & AO WFS$^c$ & 0.012\%\\
AO WFS to MEMS &  0.5\%& AO WFS & 0.019\% \\
Apodizer to MEMS &  0.25\%& Pupil viewer$^d$ or HOWFS & 0.21\%\\
Lyot Stop to MEMS & 1\%& Pupil viewer & 0.48\%\\
\hline
\end{tabular}
\end{center}
\footnotesize $^a$Tolerances are given in percentages of the corresponding pupil size.\\
$^b$ Achieved values are RSS errors in tip and tilt on the WFS and errors in x and y on the pupil viewer.\\
$^c$ Measurements in the AO WFS are in fractions of sub-apertures, with 43 sub-apertures across the pupil.\\
$^d$ The pupil is approximately 232 pupil viewer pixels across.
\normalsize
\end{table}

 In addition to the science camera in the IFS, GPI contains four other internal imaging systems:  the AO WFS, two wavefront sensors in the CAL system, and a deployable camera in the IFS assembly that can be used for imaging the IFS entrance pupil (immediately behind the Lyot stop).  The CAL wavefront sensors include a low-order IR Shack-Hartmann (LOWFS) and the high-order interferometer (HOWFS).  The interferometer can be reconfigured to block one or the other leg, so that the HOWFS camera can also act as a pupil viewer for the apodizer pupil plane.  As the APLC design is inherently chromatic\cite{sivaramakrishnan2006astrometry} and GPI operates in multiple wavelength bands, there are multiple versions of each of the three APLC components (the apodizer, FPM and Lyot stop), housed in three filter wheels at the three respective planes. Furthermore, there are two pairs of pointing and centering mirrors, located at intermediate planes not conjugate to the pupil or focal planes, which are used to steer the beam onto the AO WFS and for alignment between the CAL and IFS, respectively.  These pointing and centering pairs each have empirically derived, open loop models that account for internal flexure due to changes in temperature and the gravity vector.  For every new target, a complete system alignment is done as described in \refnum{dunnthis}, providing set points for every component. During the course of observing sequences, the open loop models are constantly active, keeping the beam path  aligned while the telescope and GPI move to track the star.
  
  \reffig{fig:gpi_schematic} shows a schematic view of the various components of GPI of interest.  In particular, the four pupil planes at the top of the schematic---the telescope pupil, the MEMS DM, the apodizer, and the Lyot stop---must all be closely aligned (as summarized in Table \ref{tbl:tolerances}) in order for the APLC to be effective. The beam must also be centered on the FPM to within 5 mas or light from the star will leak into the final image, degrading contrast.

\section{CORONAGRAPH ALIGNMENT}\label{sec:coron}
 
The basic approach to GPI's pupil alignments is discussed in \refnum{savransky2013computer}. The MEMS DM, which was precisely aligned and fixed with respect to the piezo-electric DM during GPI integration,  serves as a reference pupil plane for the entire instrument.  The telescope pupil is centered using the input fold mirror and pointing is adjusted by the tip/tilt stage, both of which run in closed loop with the AO WFS along with control loops for each of the two DMs.  The three main AO loops (controlling the two DMs and the tip/tilt stage) run full time while on-sky to maintain both pointing and a smooth wavefront, while the input fold loop (which updates at a slower rate) can be toggled at will to provide corrections for major drifts.  Offsets are also provided to the telescope's active secondary (see \refnum{rantakyrothis} for details). The beam is steered onto the CAL system by moving the AO pointing and centering mirrors while the AO loops are closed, causing the beam pointing to shift on the WFS and immediately be corrected by the tip/tilt loop, thereby changing the pointing on the CAL.  The alignment of the apodizer and Lyot pupil masks is achieved by imaging each mask independently with the IFS pupil-viewer, and comparing these images with a pattern placed on the MEMS DM to serve as a fiducial marker.  We are able to image each mask independently because each of the three filter wheels holding the APLC components includes a clear slot (or, in the case of the FPM, a fully reflective surface with no pinhole) allowing for configurations with just one APLC component at time (see \reffig{fig:pupil_ims_sky}).

\subsection{Pupil Alignment}\label{sec:pupalign}
 \begin{figure}[ht]
 \begin{center}
   \includegraphics[width=\textwidth]{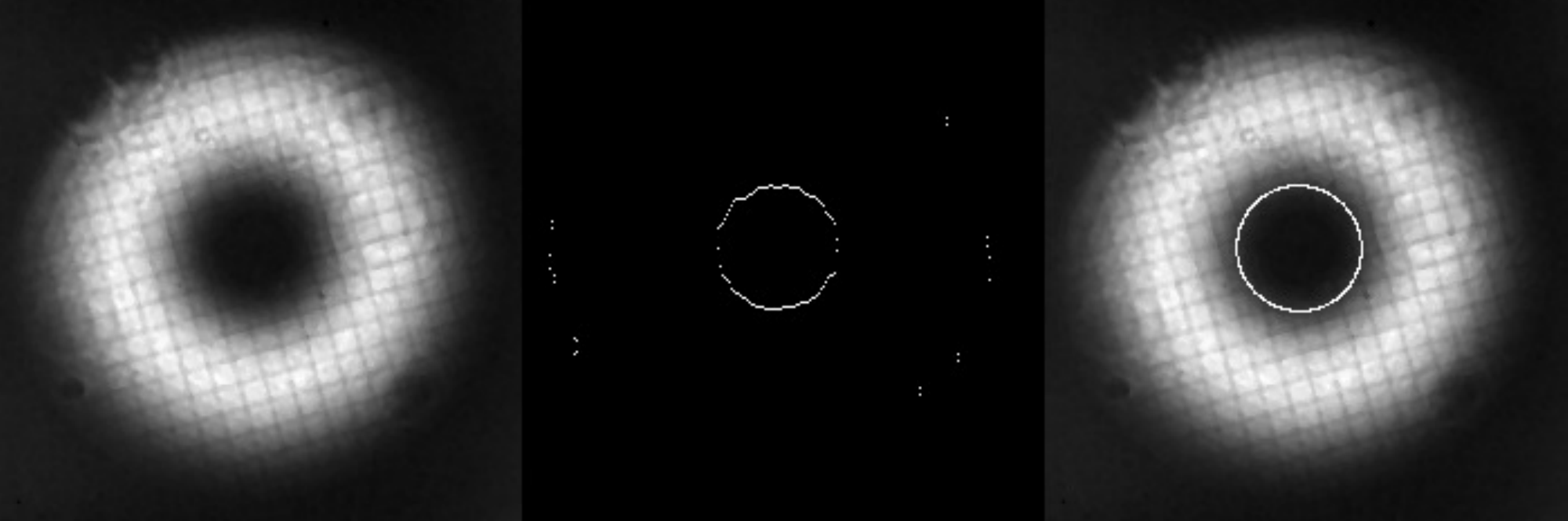} 
   \includegraphics[width=\textwidth]{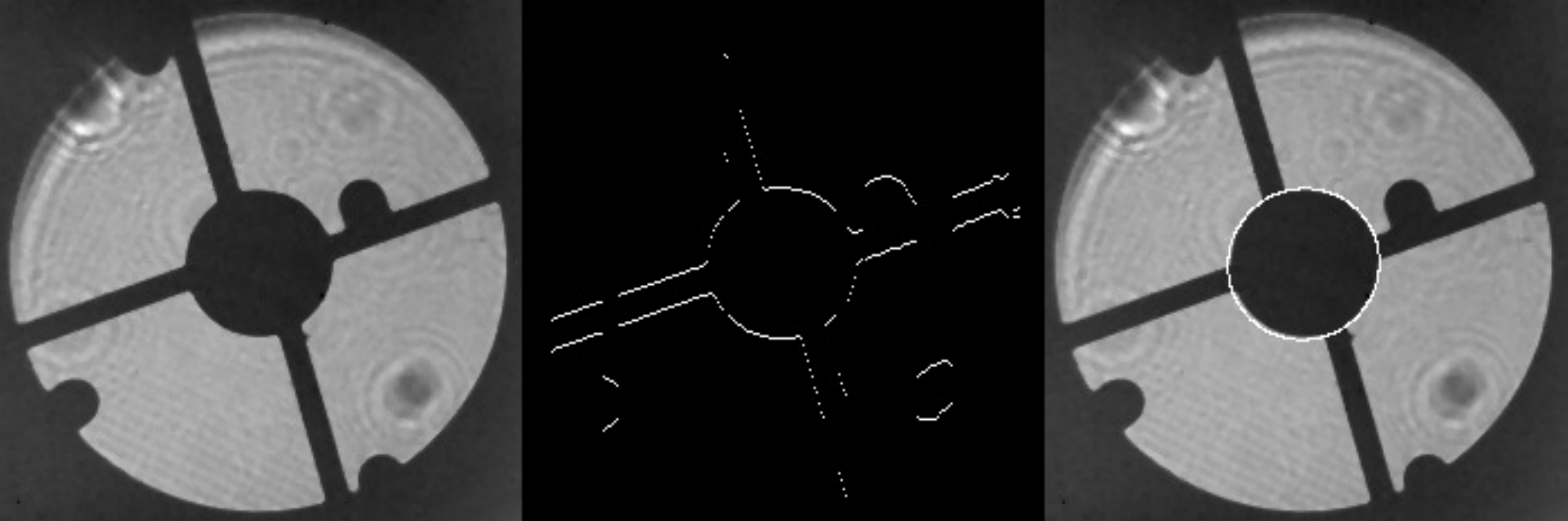}
 \end{center}
 \caption[]{ \label{fig:ellipses} Ellipse fitting to central obscuration of apodizer (top row) and Lyot stop (bottom row). \emph{Left:} Dark-subtracted, median-filtered pupil viewer image. \emph{Center:} Candidate pixels for ellipse fitting.  \emph{Right:} Best-fit ellipse overlaid on original image.}
 \end{figure}

 \begin{figure}[ht]
 \begin{center}
   \includegraphics[width=0.75\textwidth]{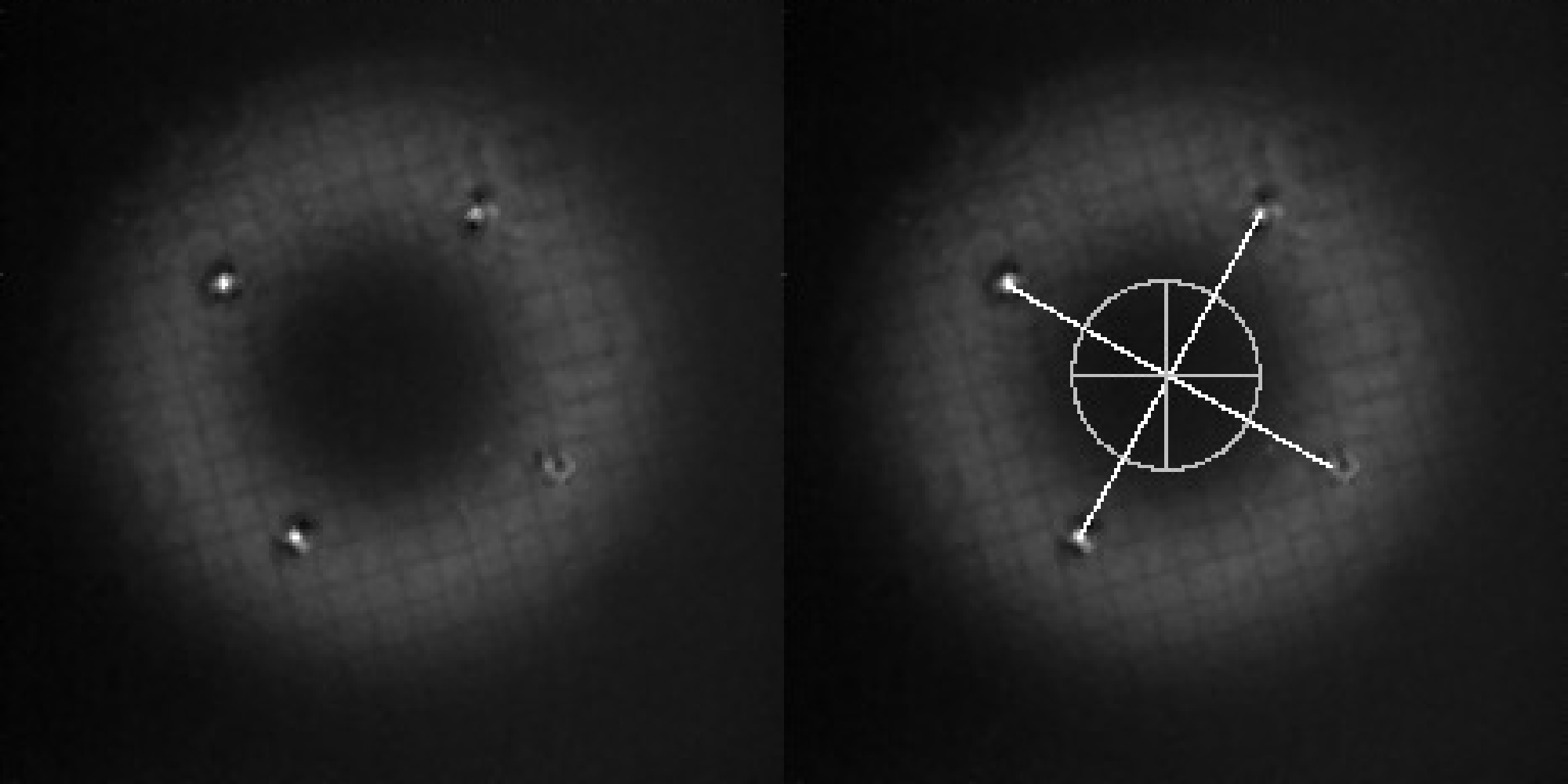} 
 \end{center}
 \caption[]{ \label{fig:apodalign} \emph{Left:} Subtracted pupil viewer images with and without the MEMS DM poke pattern. \emph{Right:} Line fits to poke pattern (white) and ellipse fit to central obscuration (gray) overlaid on original image.  In this particular case, the apodizer is misaligned from the MEMS plane by (0.99, -0.34) pupil viewer pixels, and thus the alignment would have to be updated to meet the specification in Table \ref{tbl:tolerances}.}
 \end{figure}
 
In order to identify the central point of the apodizer and Lyot pupil masks, we fit their central obscurations via an elliptical Hough transform\cite{ballard1981generalizing,xie2002new}.  While the central features of these masks are designed to be circular, various misalignment, projection, and focus effects can cause these circular shapes to become slightly deformed in the pupil images, and so we use a more generalized elliptical description.  As in \refnum{xie2002new}, we define the accumulator array only for the semi-minor axis of the ellipse being fit.  The maxima of the accumulator array represents candidate ellipse semi-minor axes and the parameters used to calculate them, so that the overall maximum can then be taken as the most likely ellipse.  This also allows us to control the range of sizes and eccentricities by rejecting proposed solutions that fall outside of some predetermined ranges. We can thus build prior knowledge (gained from a long history of pupil viewer images of each component) about the general mask characteristics into the fitting algorithm and thus significantly speed up processing.
 
The pupil viewer images are 320x240 pixels, with the IFS entrance pupil covering a square of approximately 232 pixels in height.  As another means of decreasing the fitting task time, images are pre-processed to identify a subset of candidate pixels for fitting.  A $3\times3$ pixel median filter is applied to dark subtracted images in order to remove isolated hot and cold pixels and smooth the image.  The set of candidate pixels is then identified by finding pixel locations where the difference between values of neighboring pixels is greater than a threshold set by the image median value.  The set is further filtered by calculating the slopes of each pair of pixels, and only keeping those whose slopes are a standard deviation above the mean, creating a rudimentary form of edge detection. The Hough transform is applied to only these candidate pixels, and the global maximum of the accumulator array is retained as the best-fit solution.  The transform is very robust and it is not necessary for the candidate set to include all of the pixels in the best-fit ellipse (in fact, the algorithm works when as few as 50\% of the pixels belonging to the feature of interest are identified).  It is also not necessary for the candidate pixel set to include a whole closed curve. \reffig{fig:ellipses} shows the application of the Hough transform to both the H-band apodizer (top row) as well as the corresponding Lyot stop (bottom row).  The regular grid seen in the apodizer is a microdot structure printed on each apodizer mask, which produces four satellite spots at equal distances from the true image center in the focal plane, as can be seen in \reffig{fig:fpmoff}.\cite{sivaramakrishnan2006astrometry,marois2006accurate}.  These spots allow us to precisely locate the star even when it has been blocked by the coronograph (see Refs.~\citenum{konopackythis} and \citenum{wangthis} for details).  The Lyot stops include features to block diffracted light from the telescope spiders along with tabs to cover bad MEMS DM actuators (see \reffig{fig:pupil_ims_sky}).  It is particularly interesting to note that even though a large number of pixels belonging to the Lyot spiders are identified in the edge-detection step (see bottom row of \reffig{fig:ellipses}), the algorithm has no problems finding the central obscuration.

To find the center of the reference MEMS DM plane, we simply put a symmetric phase pattern onto the flattened DM (either by `poking' individual actuators, or using sums of sine functions\cite{savransky2012focal}).  The typical pattern used produces four symmetric spots about the MEMS center in the pupil viewer image, which are then automatically detected via the box-finding algorithm described in \refnum{savransky2013computer}.  Lines are fit to the two pairs of diagonal pokes and the center of the MEMS plane is taken to be the intersection of these lines.  This is then compared with the center of the best-fit ellipse to the central obscuration of the apodizer or Lyot being aligned to determined the level of misalignment, as demonstrated in \reffig{fig:apodalign}.  In order to align the Lyot stops we make changes in centering with the CAL to IFS pointing and centering mirrors, while for the apodizers, we move the mechanism holding the masks themselves.  This is a filter wheel on a translation stage, giving us two degrees of freedom (the plane perpendicular to the beam) via coordinated rotations and translations.  In both cases, commands are calculated based on linearized transform models between the pupil viewer pixel space and the relevant mechanisms.  Commands are applied in a low frequency closed loop, with updated pupil viewer images taken in between actuations, until convergence is reached.  Convergence is determined by tracking the offsets between pupil planes in pupil-viewer pixels, with stopping points determined by the tolerances in Table \ref{tbl:tolerances} converted accordingly.  Given the high degree of repeatability in the mechanisms, and the stability of GPI overall, we have found that it is necessary to update these alignments infrequently, typically keeping a single set of default values for the entire time that GPI is on the telescope (i.e., on the order of weeks).  This is, in large part, due to the constantly running open loop models which continuously make adjustments in the pointing and centering pairs in response to changes in temperature and the gravity gradient.  The same procedure (minus the control part) is used to make daily spot checks of the mask alignments to ensure that no large drifts suddenly occur.

\subsection{Focal Plane Mask Alignment}
 
\begin{figure}[ht]
 \begin{center}
   \includegraphics[height=6.66cm,clip=true,trim=0.25in 0.5in 0.25in 0.5in]{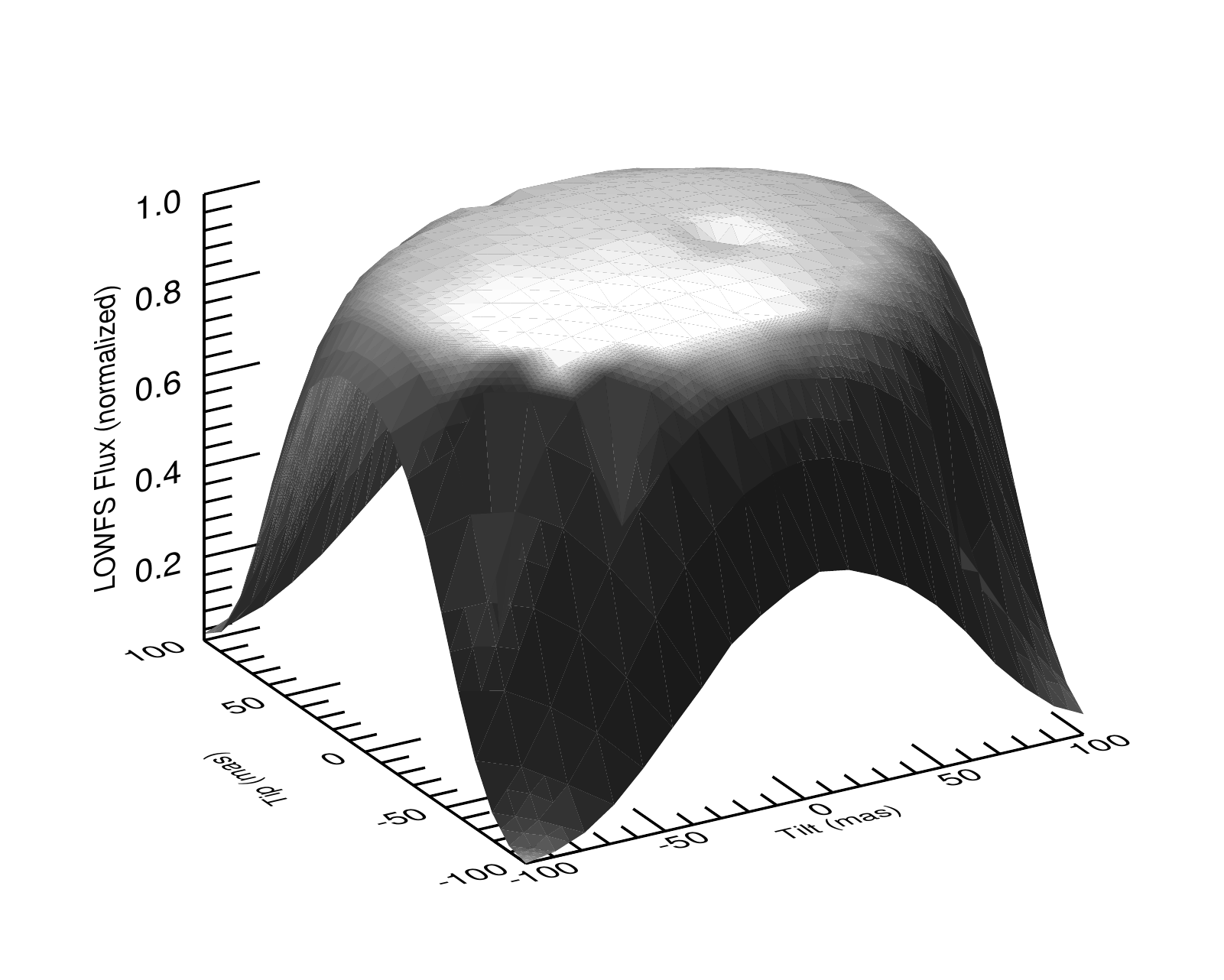} 
    \includegraphics[height=6.6cm]{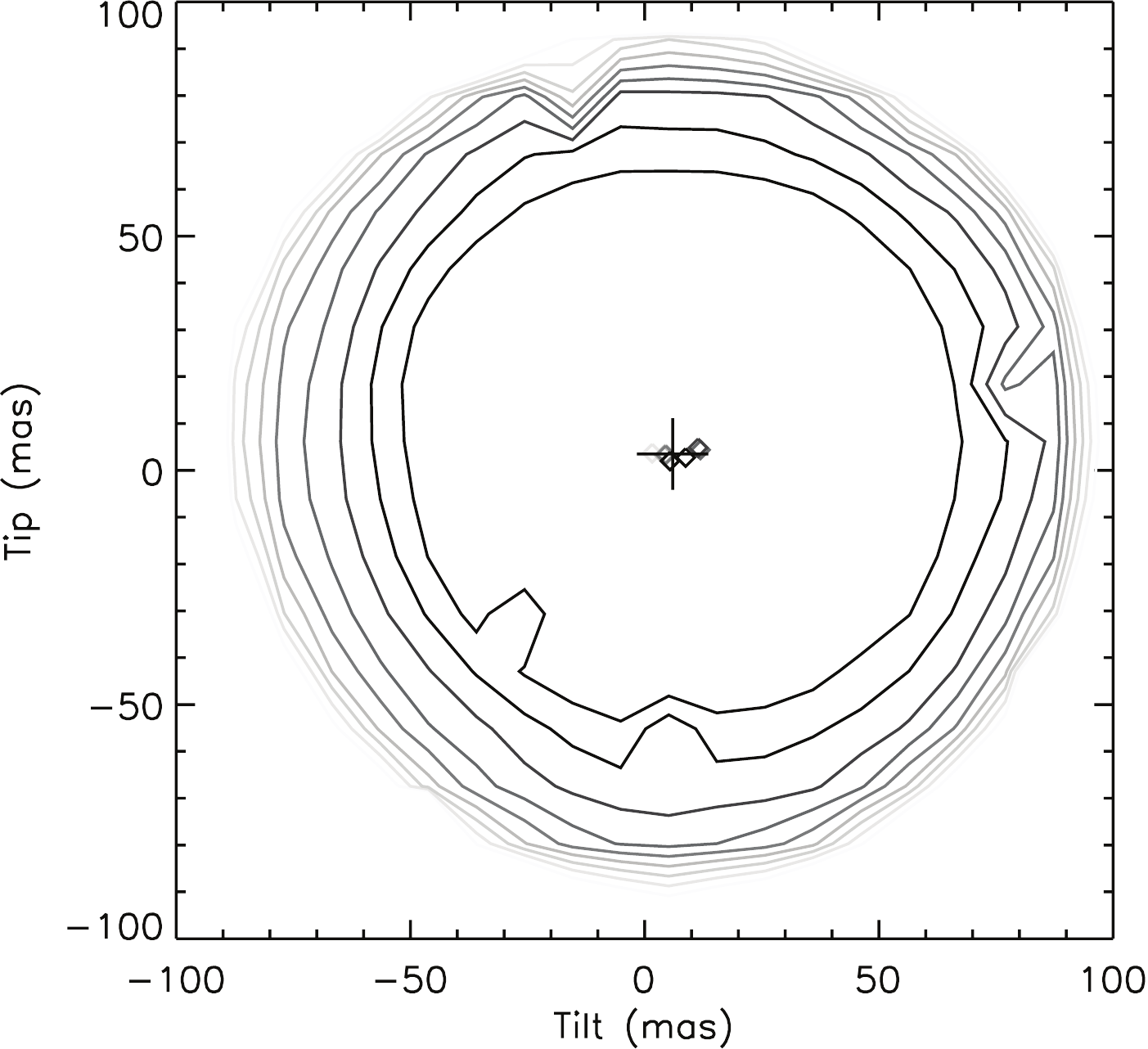}
 \end{center}
 \caption[]{ \label{fig:fpm_raster} \emph{Left:} Surface plot of raster scan of the H-band FPM with the LOWFS. \emph{Right:} Least-squares fits to contours of the raster scan, with the centers of each plot marked with diamonds.  The cross shows the mean center and error of all of the contours.  The top plateau (high flux region) represents the pinhole, with flux levels rapidly falling off as the beam moves off of the pinhole.}
 \end{figure}

GPI's focal plane mask is actually a reflecting surface with a central pinhole, which allows on-axis light to be sent to the CAL LOWFS and HOWFS.\cite{wallace2008post}  In order for the coronagraph to function properly, we must center the beam on the pinhole and hold it there throughout the course of an observing sequence, which could last for an hour or more.  The initial alignment is achieved by exploiting the symmetric nature of the focal plane mask as seen by the LOWFS.  As shown in \reffig{fig:fpm_raster}, we can map out the topography of the FPM by performing a raster scan with the AO pointing and centering mirrors with the AO loops closed.  This has the effect of moving the light on the AO WFS, which causes the AO tip/tilt loop to correct, thereby moving the beam in a controlled fashion over the FPM.  At each location, we record the total LOWFS flux, thereby generating a map of the FPM.  

The result shows a uniform plateau representing the pinhole, surrounded by a steep drop-off in flux when the beam falls onto the reflective part of the mask.  Most importantly, the slope of the flux is approximately constant at the edges of the pinhole, with a constant floor (plus some additive noise) when the light is entirely off of the pinhole.  This means that measurements taken with the beam at equal distances from the pinhole center in either direction will return the same flux value, allowing us to map the pinhole centering with LOWFS flux alone, given the ability to accurately set relative offsets in the beam location on the FPM.
 
 Using the same approach to control the location of the beam, we make symmetric excursions in two orthogonal directions (nominally the AO system's tip and tilt directions) and compare the resulting recorded LOWFS flux values.  We then move the center point of the beam in closed loop until the LOWFS fluxes to either side of the center are equal, thereby placing ourselves directly onto the pinhole.  This procedure generates a reference point for the CAL LOWFS measurement of tip and tilt.  Once complete, the CAL control loop is activated with these reference settings.  This loop uses the tip and tilt as measured by the CAL LOWFS to provide updated offsets to the AO loop (at a rate of approximately 1 Hz), which keeps the beam centered on the FPM pinhole. 

 \section{On-Sky Performance}\label{sec:onsky}
  \begin{figure}[ht]
 \begin{center}
   \includegraphics[width=0.85\textwidth]{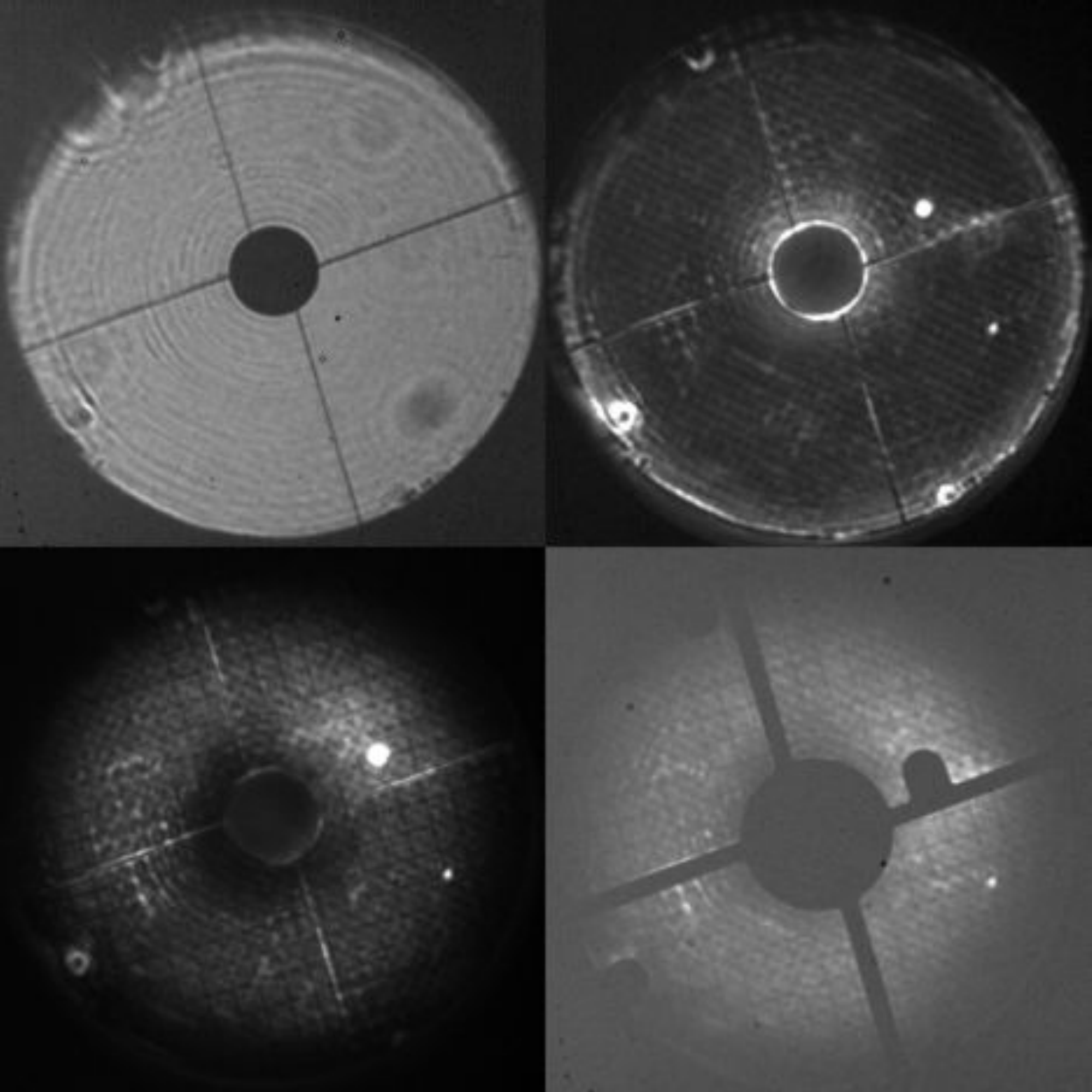} 
 \end{center}
 \caption[]{ \label{fig:pupil_ims_sky} Pupil viewer images of a 3.2 V magnitude star with all control loops closed and various configurations of the coronagraph components in place. \emph{Top Left:} No coronagraph (`direct' mode).  Gemini pupil, central obscuration and spiders are visible. \emph{Top Right:} H band focal plane mask only. Light from the star is blocked creating bright diffraction rings.  The bright spots around the edges of the pupil and in the top-right quadrant are due to bad MEMS DM actuators. \emph{Bottom Left:} Focal plane mask and apodizer only. \emph{Bottom Right:} Complete coronagraph including focal plane mask, apodizer and Lyot stop.  The regular background pattern is due to features on the MEMS.  Note that each image is individually processed and stretched for visibility. The last image contains less than 10\% of the light in the first.}
 \end{figure}
 
 While most of the alignment algorithms and operating procedures described here were designed and implemented during GPI's integration and testing period in a lab environment, we have now had ample opportunity to exercise the instrument on-sky and gauge its performance.  \reffig{fig:pupil_ims_sky} shows pupil viewer images of a 3.2 V magnitude star with all loops operating and various internal configurations of the coronagraph elements.  We can see the proper alignment of the coronagraph, with the vast majority of on-axis light removed at the final pupil plane.  As previously mentioned, we have found that GPI's various open and closed loop controllers\cite{dunnthis} are able to maintain the pupil alignments not only throughout whole observations, but for entire observing runs, with infrequent updates to mechanism default positions.  Pupil alignments performed with the internal calibration source during the daytime are repeatable to within the required tolerances and suitable for night-time, on-sky operation.  Since the Gemini pupil also has a central obscuration that can be imaged with the pupil viewer, we have found that the exact same procedure as described in \S\ref{sec:pupalign} can be used to center the telescope pupil onto the MEMS plane, as an alternative to the AO WFS measurements described above.
 
 \begin{figure}[ht]
 \begin{center}
   \includegraphics[width=0.33\textwidth,clip=true,trim=1in 0in 1in 0in]{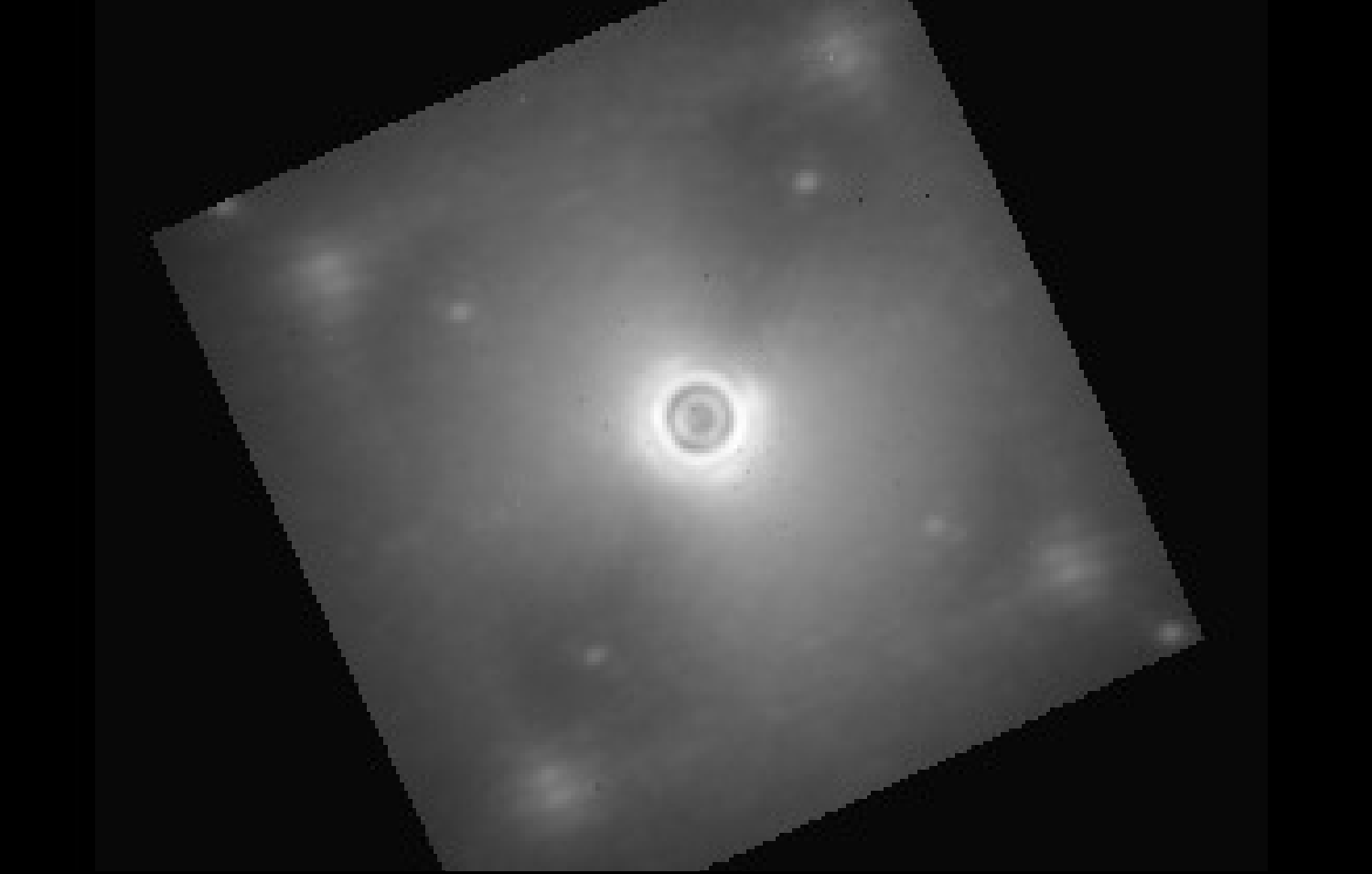} 
   \hspace{-2ex}
   \includegraphics[width=0.33\textwidth,clip=true,trim=1in 0in 1in 0in]{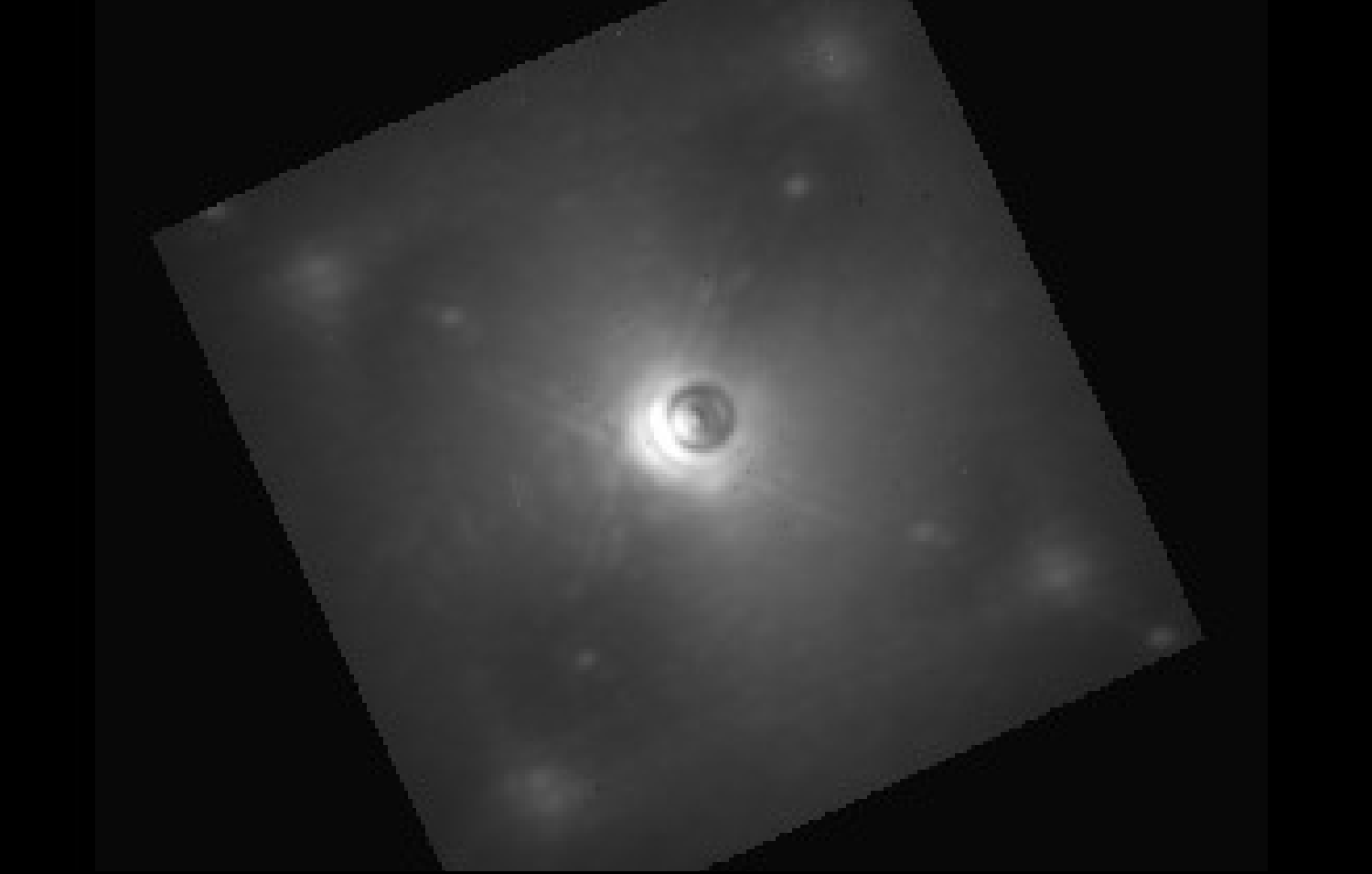} 
      \hspace{-2ex}
   \includegraphics[width=0.33\textwidth,clip=true,trim=1in 0in 1in 0in]{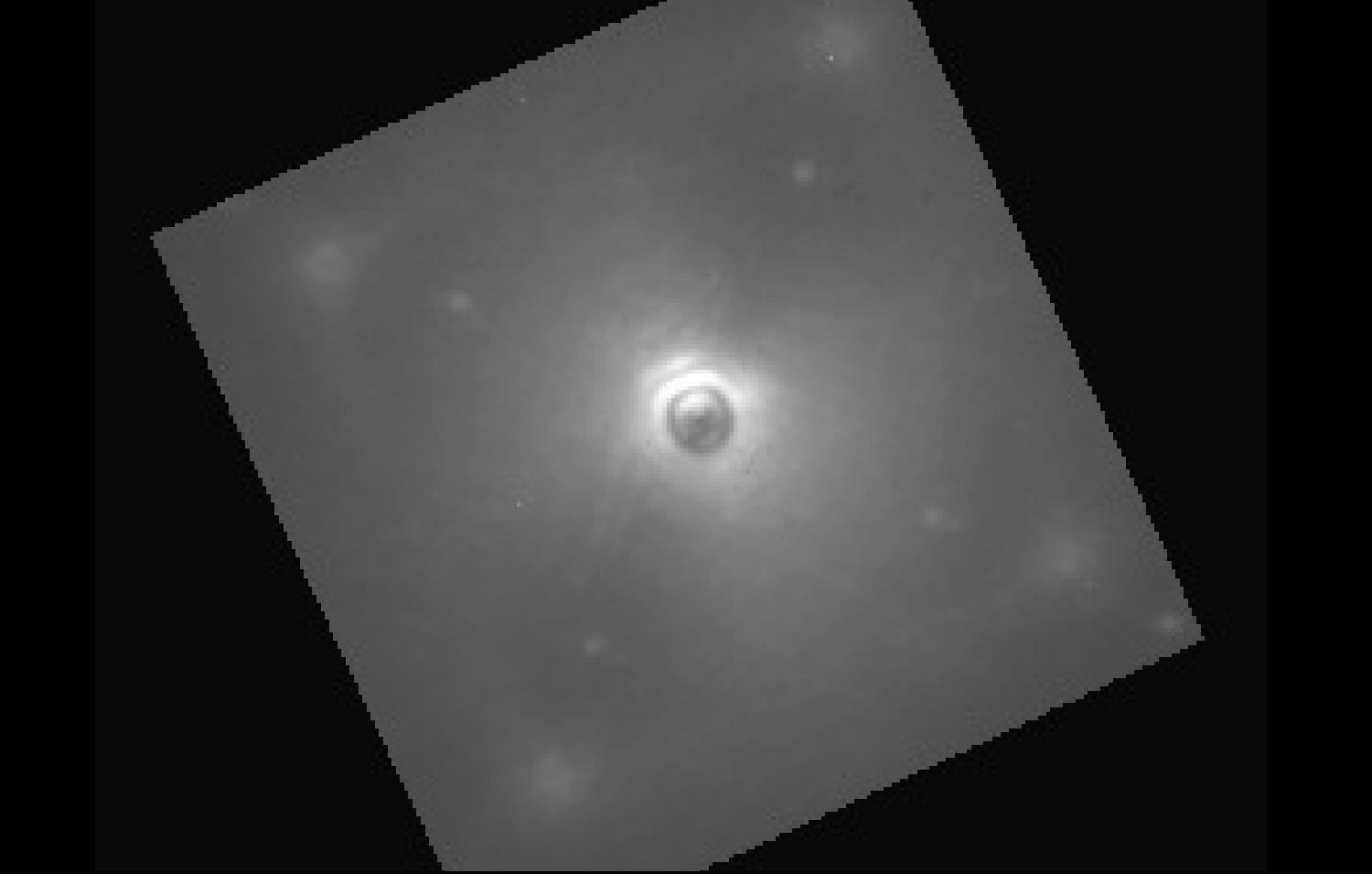} 
    \includegraphics[width=0.8\textwidth]{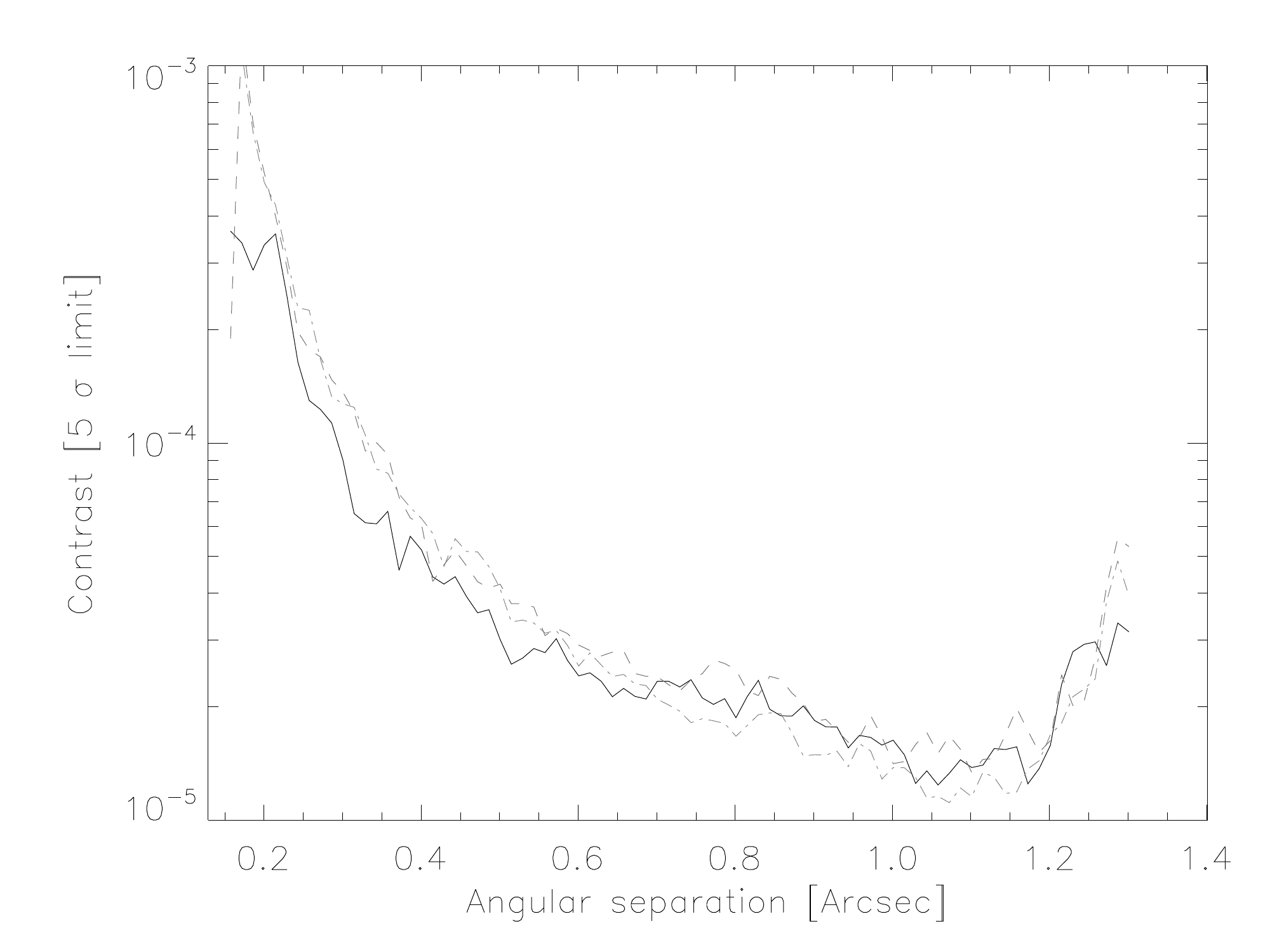} 
  \end{center}
 \caption[]{ \label{fig:fpmoff} Single slices of 60 s IFS images and their associated contrast profiles with \emph{Left:} FPM centered (solid line), \emph{Center:} FPM offset in tip (dashed line), and \emph{Right:} FPM offset in tilt (dash-dotted line).  The FPM misalignment causes a contrast degradation near the inner working angle in single frames, and raises the noise floor on whole observing sequences. Note that these are contrasts for single reconstructed cube slices with no post-processing applied other than a high-pass filter step.  They do not represent GPI's final operating contrast.}
 \end{figure}
 
 On the other hand, we have found that active maintenance of the focal plane mask alignment has a potential failure mode, especially in poor seeing conditions and when using focal plane masks with smaller pinholes (as for the Y and J spectral band coronagraphs).  Particularly large excursions in tip/tilt due to sudden turbulence or wind-shake can occasionally cause  the beam to come off the FPM pinhole causing the LOWFS to loose light.  While the LOWFS control logic contains provisions for this, if the beam swings entirely off of the pinhole then the CAL loop has no measurements that can be used to recover the beam.  Work has progressed in this area, with significant improvements in later observing runs.  If the star is significantly offset, e.g., due to uncorrected differential atmospheric refraction, we can manually center the beam onto the FPM by taking quick IFS snapshots and using our astrometric calibration\cite{konopackythis} to derive the required pointing changes.  This approach allows us to quickly bring the beam within the capture range of the CAL LOWFS loop, allowing automated operation to resume and providing coronagraph alignment stability in operating regimes beyond GPI's normal specifications.  In certain cases, where the FPM is slightly offset but light still reaches the LOWFS, we have found that we can improve alignment by manually adjusting the reference points of the CAL LOWFS loop.  We can also use this to systematically explore the effects of FPM misalignment on system performance, as demonstrated in \reffig{fig:fpmoff}, which shows the relative contrast changes for small offsets to the FPM in single IFS frames.
 
 In normal operating conditions, we have repeatedly verified that in closed loop operation, GPI's coronagraph alignment holds to the required tolerances throughout observation sequences of one hour or longer without the need for manual intervention from a human operator.  While factors such as telescope vibration have adversely affected final contrast values, these are actively being addressed and GPI's performance continues to improve (see \refnum{hartungthis} for details).  From the standpoint of reliability and robustness, GPI is already operating with a high level of automation, and is quickly nearing the point of regular facility use.
 
 \section{Conclusions and Future Work}
 We have presented a description of the methods by which we align the components of GPI's coronagraph and maintain this alignment throughout the course of whole observing sequences.  These methods are highly automated and make use of GPI's internal imaging capabilities as well as data from the science instrument.  We have also discussed our initial experiences with on-sky operations and instrument performance.  The GPI commissioning and verification team is continuously working to improve the robustness of the instrument and to complete the integration of its operations with normal Gemini South operating procedures.  GPI has already been successfully used for early science observing and continues to collect both science and engineering data.  The next stage of instrument optimization will focus on improving final contrast levels in preparation for the Gemini Planet Imager Exoplanet Survey, commencing later this year.
 
 \acknowledgements
 The authors would like to acknowledge the contributions of the entire GPI integration, commissioning and science teams and the invaluable assistance of the Gemini South telescope operations staff. GPI has been supported by Gemini Observatory, which is operated by the Association of Universities for Research in Astronomy, Inc., under a cooperative agreement with the NSF on behalf of the Gemini
partnership: the National Science Foundation (United States), the National Research
Council (Canada), CONICYT (Chile), the Australian Research Council (Australia),
Minist\'erio da Ci\'encia, Tecnologia e Inova\c{c}\=ao (Brazil), and Ministerio de Ciencia,
Tecnolog\'ia e Innovaci\'on Productiva (Argentina). Portions of this work were performed under the auspices of the U.S. Department of Energy by Lawrence Livermore National Laboratory under Contract DE- AC52-07NA27344. The GPI team makes use of Dropbox for sharing large datasets among team members and thanks Dropbox for sponsoring a team account.

\bibliography{Main,mylocalbib}   
\bibliographystyle{spiebib}

\end{document}